\documentclass[journal]{IEEEtran}
\usepackage[utf8]{inputenc}
\usepackage[english]{babel}
\usepackage{amssymb}
\usepackage{amsthm}
\usepackage{amsmath}
\usepackage{enumerate}
\usepackage{graphicx}
\usepackage{multicol}
\usepackage{multirow}
\usepackage{hyperref}
\usepackage{mathtools}
\usepackage{graphics}
\usepackage{amsfonts}
\usepackage{enumitem}
\usepackage{blindtext}
\usepackage{subfiles}
\usepackage{float}

\newcommand{\Ohm}{\Omega}

\title{ Practical Utility PV Multilevel Inverter Solutions}
\author{John Buczek and Viktor Ivankevych\\
buczek.j@northeastern.edu\\ ivankevych.v@northeastern.edu}
\begin{document}
\maketitle
\textbf{ \emph{Abstract--}Multilevel inverters are used to improve power quality and reduce component stresses. This paper describes and compares two multilevel cascaded three phase inverter implementations with two different modulation techniques: Phase Shifted Pulse Width Modulation, and Nearest Level Control. Further analysis will show required number of inverter levels with respect to modulation techniques to provide desired power and power quality to resistive load or grid. Cascaded inverter will be designed and simulated to draw power from PV cells.}
\section{Introduction}
\hspace{16pt} A multi-level inverter is a power electronic system that synthesizes a desired voltage output from several levels of DC voltages as inputs \cite{pv-citation-1}. Today, there are many different topologies of multilevel converters including, but not limited to, Diode-Clamped, Flying Capacitor, and Cascade H-bridge (CHB). While the topologies may be different, they all offer similar beneficial features. For sinusoidal outputs, multilevel converters improve their output voltage in quality as the number of levels of the converter increase, thus decreasing the Total Harmonic Distortion (THD) \cite{Kouro}. For this reason and others, multilevel converters have been used for high power photovoltaic (PV) inversion, electric motor drivers in electric vehicles, and other research and commercial applications \cite{Kouro, ev-citation-1,pv-citation-2, ev-citation-1, Franquelo}. Although, technological problems such as reliability, efficiency, the increase of the control complexity, and the design of simple modulation methods have slowed down the application of multilevel converters \cite{Kouro}. 
\begin{figure}[h]
    \centering
    \includegraphics[height=9cm]{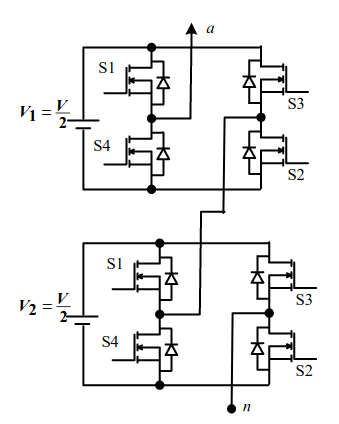}
    \caption{5 Level Cascade H-bridge Converter \cite{pv-citation-2}}
    \label{chb-5level}
\end{figure}
\\
\par \hspace{16pt} Figure \ref{chb-5level} shows a 5 level CHB converter. As can be seen, CHB converters consist of multiple MOSFET (or equivalent) H-bridges that are connected in series. Each H-bridge having its own isolated DC voltage source. For the shown 5 level case, it requires two H-bridges that can be configured to output the 5 levels: +V, +V/2, 0, -V/2, and -V. While the theory of adding and subtracting isolated voltage sources is simple, such is harder to do in practice. Common methods include using isolated DC-DC converters such as flyback and forward converters that have transformers with multiple secondary windings \cite{Kouro}. Others use individual, or multiple isolated sets of PVs that power individual H-bridges \cite{pv-citation-1, pv-citation-2}. 
\\
\par \hspace{16pt} This paper uses two of the common modulation techniques for CHB converters, Phase Shifted PWM (PSPWM) and Nearest Level Control (NLS), to propose designs for a utility 3 phase PV inverter. Our designs will display many of the common advantages and disadvantages of the different modulation techniques for CHB. Our design requirements were to develop a 3 phase utility PV CHB inverter to supply 125kW  at $480 V_{L-L}^{RMS}$ with a THD below 5\%.

\section{Nearest Level Switching}
    \par \hspace{16pt}
    The Nearest-Level Switching control for a multilevel converter compares a control sine wave to DC voltage levels. For a Cascading Multi-level Converter consisting of \emph{N} H-bridges, each trigger when the voltage control sinusoidal is greater than their respective threshold given by the equation:
    \begin{align}
        V_{thresh}(i) = V_{pk} \frac{2i-1}{2N} \label{nls-vthresh}
    \end{align}
    \par \hspace{16pt} Where \emph{N} is the number of H-bridges equal to $N = \frac{L-1}{2}$ for an \emph{L} level Cascade H-bridge (CHB), and \emph{i} is the switch number which ranges from 0 to \emph{N-1}. 
    \begin{figure}[ht]
        \centering
        \includegraphics[width=0.5\textwidth]{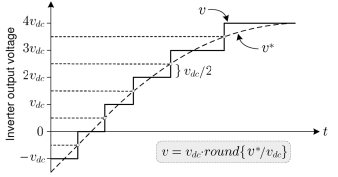}
        \caption{Nearest Level Switching Waveform Synthesis \cite{Kouro}}
        \label{nls-theory-wave}
    \end{figure}
    \par \hspace{16pt} Figure \ref{nls-theory-wave} shows the waveform for the Nearest Level Switching. It should be noted that for every DC voltage source, the threshold voltage is at half of the DC value. Additionally, the peak output voltage is equal to $N*V_{DC}$. 
    
    \begin{figure}[ht]
        \centering
        \includegraphics[width=0.5\textwidth]{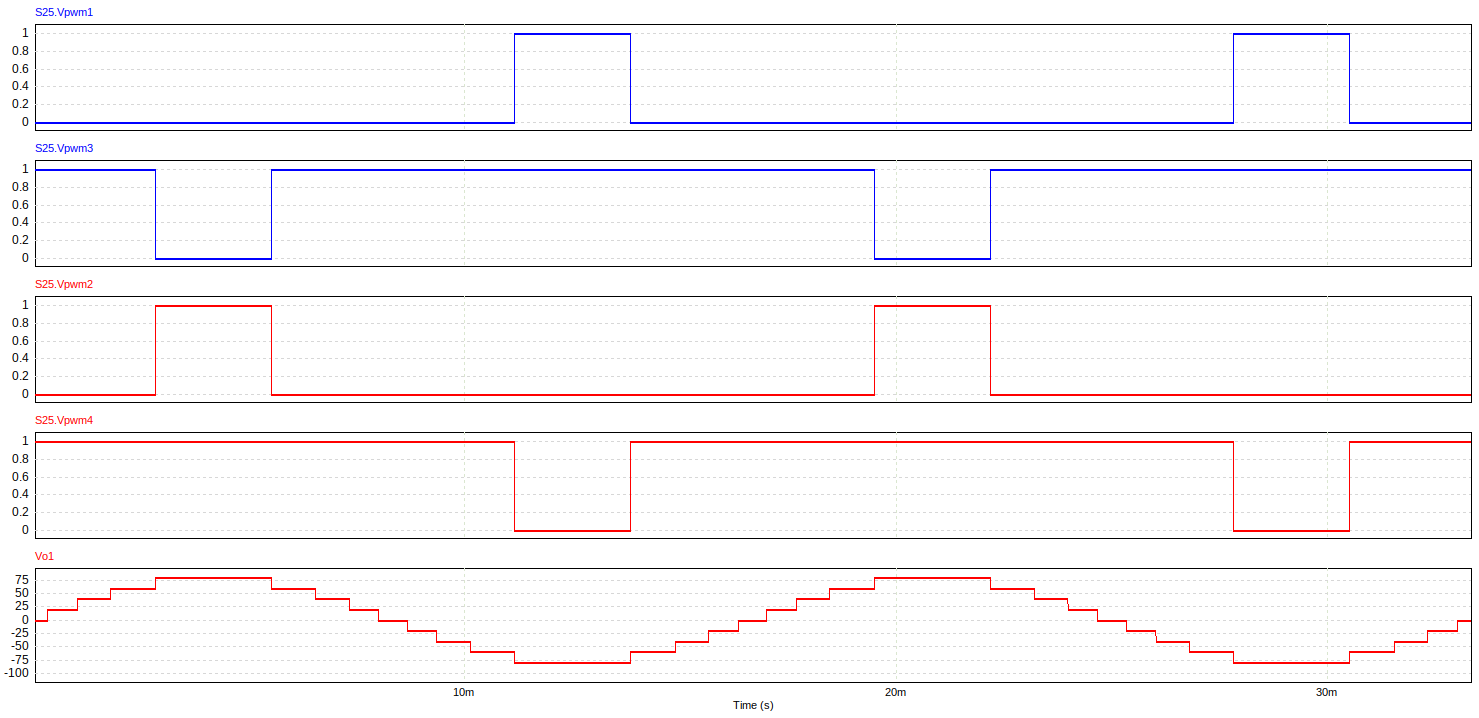}
        \caption{9-Level NLS CHB PWM Waveforms}
        \label{nls-pwm-wave}
    \end{figure}
    \par \hspace{16pt} Figure \ref{nls-pwm-wave} shows the PWM waveforms for the top H-bridge of a 9-level NLS CHB. The PWM for switches 1 and 3 are not the same, as is the case for switches 2 and 4. For a single H-bridge inverter, the diagonal switches could have the same PWM input, but that is not possible for multilevel converters. In order to achieve all of the voltage steps, either voltages need to be added and subtracted, or the outputs of H-bridges need to be shorted \cite{Franquelo}. Our approach was to use the shorting technique. This essentially adds an equivalent of a zero state in Space Vector Modulation \cite{Kouro}. This output shorting was implemented by making the PWM for switch 3 the inverse of PWM 2 and PWM 4 the inverse of PWM 1. This shorted the outputs through switches 4 and 3. In practice, both zero states of switches 3,4 as well as switches 1,2 should be used in order to have more even wear on individual switches. 
    
\section{Nearest Level Switching Resistive Load RMS and THD Calculations}
    \par \hspace{16pt}For a cascading H-Bridge Multilevel Inverter with $L$ levels, using the Nearest-Level-Switching technique, the switching point $\alpha_i$ for level $i = 0$ to $i=\frac{L-1}{2}$, are given by the equation:
    \begin{align}
        \alpha_i = sin^{-1}\left( \frac{2i+1}{L-1}\right)   \label{eqn-alpha}
    \end{align}
    \subsection{Root Mean Squared}
        \par \hspace{16pt} From \cite{Mohan}, the equation for calculating the RMS ($X^{RMS}$) of a function $x(t)$ is given as:
        \begin{align}
            X^{RMS} = \sqrt{\frac{1}{T}\int_0^T (x(t))^2 dt} \label{RMS_equation}
        \end{align}
        
        \subsubsection{3-Level}
                For the simple 3-level Inverter with NLS, the RMS voltage of a resistive load can be calculated to be:
                \begin{align*}
                    V_{3-L}^{RMS} & = \sqrt{\frac{1}{\pi}\int_{\alpha_0}^{\pi-\alpha_0}(V_m)^2 dt}\\
                    & = V_m \sqrt{ \frac{1}{\pi}\left[ t \right]_{\alpha_0}^{\pi-\alpha_0}}\\
                    & = V_m \sqrt{\frac{1}{\pi}\left( \pi - \alpha_0 - \alpha_0 \right)}\\
                    & = V_m \sqrt{1 - \frac{2}{\pi}\alpha_0}
                \end{align*}
                \par \hspace{16pt} For $\alpha_0 = sin^{-1}\left( \frac{1}{2}\right) = \frac{\pi}{6}$
                \begin{align*}
                    V_{3-L}^{RMS} & = V_m \sqrt{\frac{2}{3}} 
                \end{align*}
            \subsection{5-Level} For the simple 5-level Inverter with NLS, the RMS voltage of a resistive load can be calculated to be:
                \begin{align*}
                    V_{5-L}^{RMS} & = \biggl\{\frac{1}{\pi}\biggl( \int_{\alpha_0}^{\alpha_1}\biggl( \frac{V_m}{2}\biggl)^2 dt + \int_{\alpha_1}^{\pi - \alpha_1}\biggl( V_m\biggl)^2 dt \\
                    &+ \int_{\pi-\alpha_1}^{\pi-\alpha_0}\biggl( \frac{V_m}{2}\biggl)^2 dt\biggl)\biggr\}^{1/2}\\ \\
                    & = V_m \sqrt{\frac{1}{\pi}\left( \left[\frac{t}{4}\right]_{\alpha_0}^{\alpha_1} + \left[t\right]_{\alpha_1}^{\pi - \alpha_1}  + \left[\frac{t}{4}\right]_{\pi -\alpha_1}^{\pi -\alpha_0}\right)}\\
                    & = V_m \sqrt{\frac{1}{\pi} \left( \pi - \frac{1}{2}\alpha_0 - \frac{3}{2}\alpha_1 \right)}\\
                    & = V_m \sqrt{1 - \frac{1}{2 \pi}\alpha_0 + \frac{3}{2 \pi} \alpha_1}
                \end{align*}
                \par \hspace{16pt} For $\alpha_0 = sin^{-1}\left( \frac{1}{4}\right)$ and $\alpha_1 = sin^{-1}\left( \frac{3}{4}\right)$
                \begin{align*}
                    V_{5-L}^{RMS} & \approx V_m 0.7449 
                \end{align*}
            \subsection{7-Level}
                 For the simple 7-level Inverter with NLS, the RMS voltage of a resistive load can be calculated to be:
                \begin{align*}
                    V_{7-L}^{RMS} & = \biggl\{\frac{1}{\pi}\biggl( \int_{\alpha_0}^{\alpha_1}\left( \frac{V_m}{3}\right)^2 dt + \int_{\alpha_1}^{\alpha_2}\left( \frac{2 V_m}{3}\right)^2 dt \\
                    &+ \int_{\alpha_2}^{\pi - \alpha_2}\left( V_m\right)^2 dt + \int_{\pi - \alpha_2}^{\pi - \alpha_1}\left( \frac{2V_m}{3}\right)^2 dt  \\
                    &+ \int_{\pi-\alpha_1}^{\pi-\alpha_0}\left( \frac{V_m}{3}\right)^2 dt\biggl)\biggl\}^{1/2}\\ 
                    \\
                    & = V_m \biggl\{\frac{1}{\pi} \biggl( \left[\frac{t}{3} \right]_{\alpha_0}^{\alpha_1} + \left[\frac{2t}{3} \right]_{\alpha_1}^{\alpha_2}  \\
                    &+ \left[t \right]_{\pi-\alpha_2}^{\alpha_2} + \left[\frac{2t}{3} \right]_{\pi-\alpha_1}^{\pi-\alpha_2} + \left[\frac{t}{3} \right]_{\pi-\alpha_1}^{\pi-\alpha_0}\biggl)\biggl\}^{1/2}\\
                    \\
                    & = V_m \sqrt{\frac{1}{\pi}\left(\pi - \frac{2}{9}\alpha_0 - \frac{6}{9}\alpha_1 - \frac{10}{9}\alpha_2 \right)}\\
                    & = V_m \sqrt{1 - \frac{2}{9\pi}\alpha_0 - \frac{6}{9\pi}\alpha_1 - \frac{10}{9\pi}\alpha_2}
                \end{align*}
                \par \hspace{16pt} For $\alpha_0 = sin^{-1}\left( \frac{1}{6}\right)$, $\alpha_1 = sin^{-1}\left( \frac{3}{6}\right)$, and $\alpha_2 = sin^{-1}\left( \frac{5}{6}\right)$
                \begin{align*}
                    V_{7-L}^{RMS} & \approx V_m 0.7217 
                \end{align*}
            \subsection{L-Level}
                 From the previous derivations and Equation \ref{eqn-alpha}, a pattern for the RMS voltage can be seen. For a multilevel cascade H-Bridge Inverter with L levels, the equation for the RMS voltage of a resistive load can be given by:
                \begin{align}
                    V_L^{RMS} &= V_m \sqrt{1 - \sum^{N-1}_{i=0}\left( \frac{2(2i+1)}{\pi*N^2}*\alpha_i\right)} \nonumber \\
                    &= V_m \sqrt{1 - \frac{2}{\pi N^2} \sum^{N-1}_{i=0}\left( sin^{-1}\left(\frac{2 i + 1}{L -1} \right) \left(2i+1 \right) \right)} \label{eqn-L-level-RMS}
                \end{align}

    \subsection{Fourier Series Expansion}
        \par \hspace{16pt} The Fourier Series Expansion was also performed on the ideal cascade H-bridge Inverter. In the ideal case with a purely resistive load, the output waveform has odd symmetry and quarter-wavelength symmetry inherently. From \cite{Mohan}, the Fourier Series Expansion Coefficients for such symmetry are given as:
        \begin{align*}
            a_h &= 0\\
            b_h &= \frac{4}{\pi}\int_0^{\frac{\pi}{2}}\left( x(t) sin(h\omega t )\right) d\omega t
        \end{align*}
        
            \subsubsection{3-Level}
                 For the simple 3-level Inverter with NLS, the Fourier Series Expansion Coefficients of a resistive load can be expressed as:
                \begin{align*}
                    b_h (3) & = \frac{4}{\pi}\int_{\alpha_0}^{\frac{\pi}{2}}\left( V_m sin(h\omega t )\right) d\omega t\\
                    & = \frac{4V_m}{\pi h} \left[-cos(h \omega t) \right]^{\frac{\pi}{2}}_{\alpha_0}\\
                    & = \frac{4V_m}{\pi h} cos(h \alpha_0)
                \end{align*}
            \subsection{5-Level}
                 For the simple 3-level Inverter with NLS, the Fourier Series Expansion Coefficients of a resistive load can be expressed as:
                \begin{align*}
                    b_h (5) & = \frac{4}{\pi} \biggl(\int_{\alpha_0}^{\frac{\pi}{2}}\left( \frac{V_m}{2} sin(h\omega t )\right) d\omega t \\
                    &+ \int_{\alpha_1}^{\frac{\pi}{2}}\left( \frac{V_m}{2} sin(h\omega t )\right) d\omega t \biggl)\\
                    \\
                    & = \frac{2 V_m }{\pi h}\left( \left[-cos(h\omega t) \right]_{\alpha_0}^{\frac{\pi}{2}} + \left[-cos(h\omega t) \right]_{\alpha_1}^{\frac{\pi}{2}} \right)\\
                    & = \frac{2 V_m }{\pi h}\left(  cos(h\alpha_0)  +cos(h\alpha_1) \right)\\
                    & = \frac{2 V_m }{\pi h}\left(cos(h\alpha_0) +  cos(h\alpha_1) \right)
                \end{align*}
            \subsection{L-Level}
                 For a multilevel cascade H-Bridge Inverter with L levels, the Fourier Series Expansion Coefficients of a resistive load can be expressed as:
                \begin{align*}
                    b_h (L) & = \frac{4}{\pi} \left(\sum^{N-1}_{i=0} \int_{\alpha_i}^{\frac{\pi}{2}}\left( \frac{V_m}{N} sin(h\omega t )\right) d\omega t  \right)\\
                    & = \frac{4 V_m}{\pi h N} \sum^{N-1}_{i=0} \left( cos \left(h \alpha_i\right) \right)
                \end{align*}
                \par \hspace{16pt} This can be simplified using the trigonometry identity:
                \begin{align*}
                    cos(sin^{-1}(x)) = \sqrt{1 - x^2}
                \end{align*}
                \par \hspace{16pt} From Equation \ref{eqn-alpha}, the Fourier Series first Coefficient can be simplified to be:
                \begin{align}
                    b_1 (L) & =\frac{4 V_m}{\pi N} \sum^{N-1}_{i=0} \left( \sqrt{1 - \left( \frac{(2i+1)}{L-1} \right)^2}\right) \label{eqn-fourier}
                \end{align}
        
    \subsection{Total Harmonic Distortion}
        \par \hspace{16pt} From \cite{Mohan}, the equation for the Total Harmonic Distortion (THD) is given by:
        \begin{align}
            THD = \frac{\sqrt{(X^{RMS})^2 - (X^{RMS}_1)^2}}{X^{RMS}_1} \label{eqn-thd}
        \end{align}
        \par \hspace{16pt} From Equation \ref{eqn-fourier}, the equation for the RMS magnitude of the first harmonic can be calculated as:
        \begin{align}
            V^{RMS}_1 (L) = \frac{4 V_m}{\pi N \sqrt{2}} \sum^{N-1}_{i=0} \left( \sqrt{1 - \left( \frac{(2i+1)}{L-1} \right)^2}\right) \label{eqn-first-rms}
        \end{align}
        \par \hspace{16pt} Equations \ref{eqn-first-rms}, \ref{eqn-L-level-RMS}, and \ref{eqn-thd} were combined and plotted using python (Appendix A) to calculate the THD  of an L-level CHB inverter with a resistive load. At the same time, PSIM simulations for the same number of levels were run with resistive loads and compared against one another.

        \begin{table}[h]
            \centering
            \caption{PSIM Simulated and Theoretically Calculated THD}
            \begin{tabular}{ |c|c|c|} 
                            \hline
                            Levels (L) & PSIM THD (\%) & Calculated THD(\%)\\
                            \hline
                                3&	31.0512	&31.08419\\
                                5&	17.5799	&17.6012\\
                                7&	12.2126	&12.2272\\
                                9&	9.35322	&9.363669\\
                                11&	7.58321	&7.587252\\
                                13&	6.3712	&6.378124\\
                                15&	5.49467	&5.502021\\
                                17&	4.83621	&4.837995\\
                                19&	4.31314	&4.317328\\
                                21&	3.89612	&3.89809\\
                                23&	3.55342	&3.553263\\
                                25&	3.26193	&3.264629\\
                                27&	3.017	&3.01947\\
                             \hline
                \end{tabular}
            \label{thd-table}
        \end{table}

        \par \hspace{16pt} Table \ref{thd-table} shows the THD results from both the Theoretical Calculated THD and the PSIM simulated THD for 3 levels to 27 levels. The two data sets are consistent with each other. From the Table, as the number of levels increases, the THD decreases. This quantitatively confirms that the output sinusoidal quality improves with more and more additional levels of the CHB.   

\section{Nearest Level Switching Simulation Design}
    \par \hspace{16pt} As previously mentioned, the design goals were to produce a 60 Hz 3 phase inverter at $480 V_{L-L}^{RMS}$ with a real power output of 125 kW. The THD of our inverter also needed to be below 5 \%. These specifications were consistent with other commercially available PV inverters \cite{industry-inverters}. 
    \begin{figure}[H]
        \centering
        \includegraphics[width=0.5\textwidth]{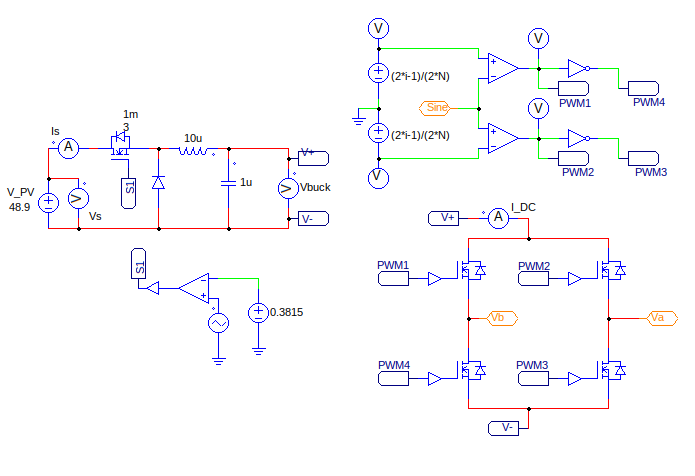}
        \caption{Nearest Level Switching Simulated H-bridge and Buck Converter}
        \label{nls-h-and-buck}
    \end{figure}
    \par \hspace{16pt} Figure \ref{nls-h-and-buck} shows the PSIM simulation for the NLS-CHB inverter individual H-bridge and buck converter. From the previous section, it was found that for a resistive load, the theoretical THD of a 27 level NLS-CHB was approximately 3\%, so a 27 level inverter was chosen. From the number of levels and the peak grid voltage of $480 V_{L-L}^{RMS}$, the individual H-bridge voltage was calculated to be $V_{dc} = 480\sqrt{2/3}/13\approx30.15V$. Our simulations did not include PV maximum power point tracking, so our simulations used a constant voltage source equal to the maximum voltage output of a single PV from the datasheet in \cite{solar-panels}. \\
    \par \hspace{16pt}From \cite{Krein}, for a buck converter, we know:
    \begin{align}
        V_o & = DVs \label{buck-d}\\
        \Delta i_L & \approx \frac{V_l* \Delta t}{L}\nonumber\\
        & \approx \frac{D V_s (1-D)T}{L}\nonumber\\
        L & \approx \frac{D V_s (1-D)T}{\Delta i_L} \label{buck-L}\\
        \Delta V_c & \approx \frac{T \Delta i_L}{8C} \label{buck-C}
    \end{align}
    \par \hspace{16pt} We expect $I_{Lpk}\approx30A$. Let $f_s = 200kHz$, $V_s = 48.9V$, $V_o = 480\sqrt{2/3}/13\approx30.15V$,$\Delta V_c = 4V$ , and $\Delta i_L = 5\% = 6A$. Then:
    \begin{align*}
        D &= 0.6165\\
        L &= 9.633 \mu H \approx 10 \mu H\\
        C &= 937 nF \approx 1 \mu F
    \end{align*}
    \par \hspace{16pt} The capacitor and inductor were intentionally kept small, but above critical values such to reduce their equivalent impedance when placed in series. Once the buck converters were designed, a single phase was simulated against a grid ac voltage source with a phase shift of -2.5 degrees in order to determine the desired cutoff frequency for the output filter. From our simulations, a cutoff frequency of approximately 700Hz was selected based on the output current harmonics. A simple LC low pass filter was chosen, with a cutoff frequency given by \cite{Krein} as: 
    \begin{align}
        f_c = \frac{1}{2 \pi \sqrt{LC}} \label{f-c-LC}
    \end{align}
    \par \hspace{16pt} An inductor $L_f = 1mH$ and a capacitor $C_f=50\mu F$ were selected.  
\section{Nearest Level Switching Simulation Simulation}
    \subsection{NLS Results}
    \par \hspace{16pt} The NLS-CHB was first simulated with a single phase using ideal switches and a phase angle of -2.5 degrees from the grid voltage source. The simulation was able to reach a power level of 8.5 kW per phase at the desired line to neutral voltage level. The output current RMS was 30.66A and the current THD was 3.02\%. At this current level, using the PV arrays from \cite{solar-panels}, we would need two in parallel going connected to each buck converter. 
    \\
    \par \hspace{16pt} The NLS-CHB was then simulated with a single phase using the PSIM default lossy switching models for NMOS MOSFETs. In addition, all reactive components were given series resistance values of 50m $\Ohm$. The circuit was simulated for 0.5s and the output THD increased to 4.486\% at a power factor of 99.56\%. 
    
    \begin{figure}[ht]
        \centering
        \includegraphics[width=0.5\textwidth]{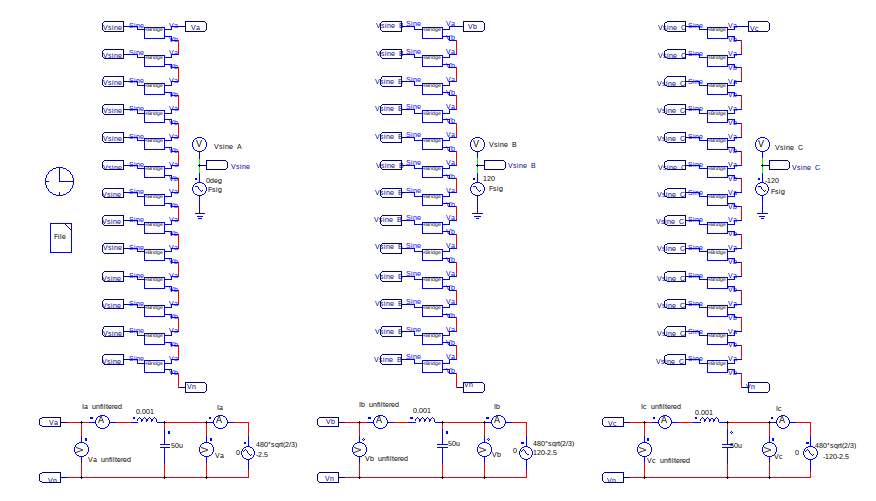}
        \caption{Nearest Level Switching PSIM 3-Phases (Y Connection)}
        \label{nls-3phase}
    \end{figure}
    
     \begin{figure}[ht]
        \centering
        \includegraphics[width=0.5\textwidth]{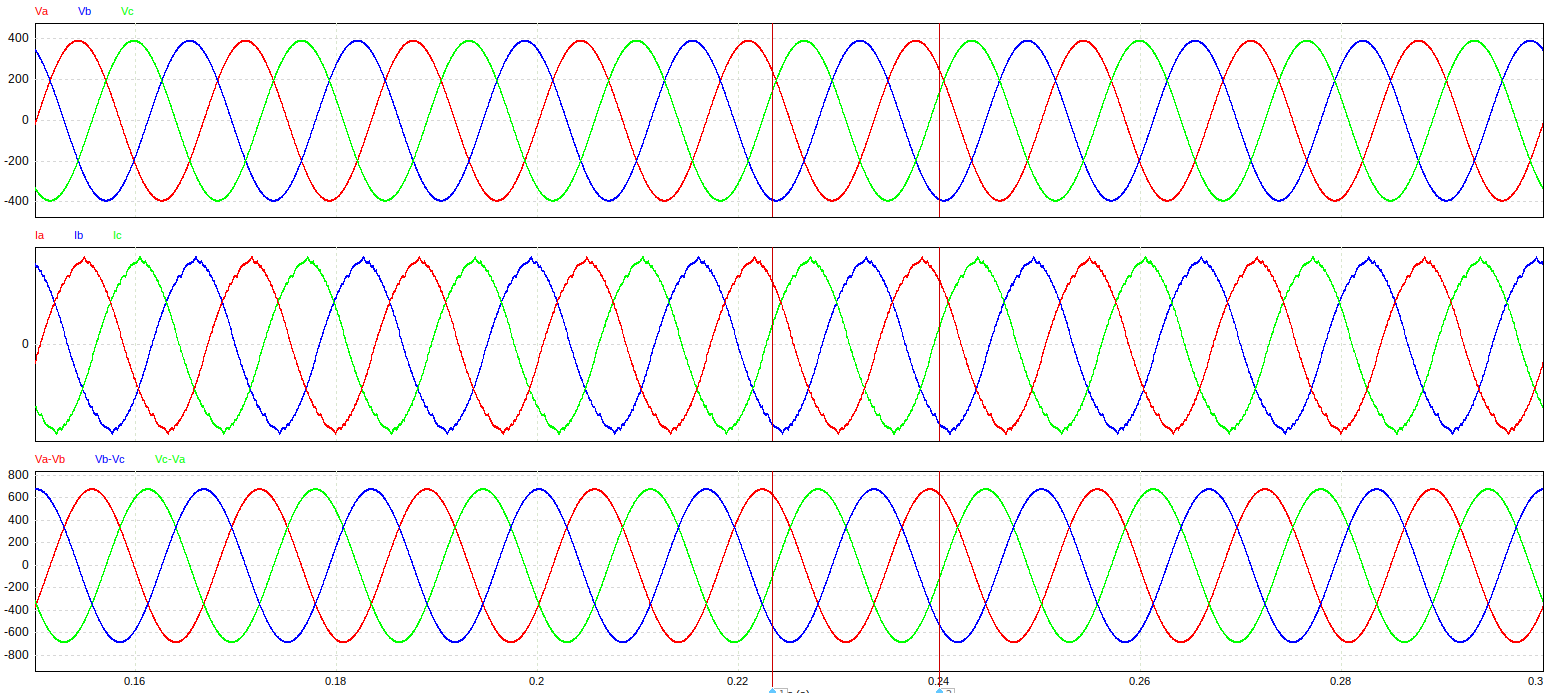}
        \caption{Nearest Level Switching 3 Phase Voltage and Current Waveforms}
        \label{nls-3phase-wave}
    \end{figure}
    
     \begin{figure}[ht]
        \centering
        \includegraphics[width=0.5\textwidth]{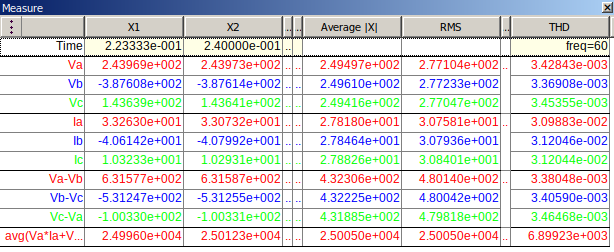}
        \caption{Nearest Level Switching 3 Phase Voltage and Current Characteristics}
        \label{nls-3phase-psim}
    \end{figure}
    \par \hspace{16pt} The NLS-CHB was finally simulated with all three phases, each with a respective phase shift of -2.5 degrees. Figure \ref{nls-3phase} shows the PSIM file of the NLS-CHB three phase simulation. Figures \ref{nls-3phase-wave} and \ref{nls-3phase-psim} show the output waveforms and characteristics respectively. The desired output voltage of $480 V_{L-L}^{RMS}$ was achieved. The output real power was approximately 25kW with an output current THD of approximately 3.12\%. From our simulation, in order to achieve our design requirements, 5 of such three phase inverters would need to be placed in parallel. 
    \begin{figure}[ht]
        \centering
        \includegraphics[width=0.5\textwidth]{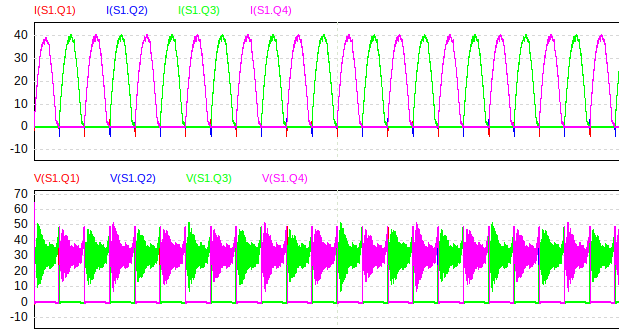}
        \caption{Nearest Level Switching H-bridge Switch Stresses}
        \label{nls-switch-stresses}
    \end{figure}
    \par \hspace{16pt} Figure \ref{nls-switch-stresses} displays the individual switch stresses for the H-Bridge MOSFETs. Each MOSFET in each H-bridge needed to conduct a peak of 50A and block a peak of 40V. Base on design experience, voltage and current ratings are desired to be increased by ~ 150 \% to maintain safe operation at extreme performance cases. For a $480 V_{L-L}^{RMS}$ system, these are comparatively small values \cite{Franquelo}. While in this topology there are more switches, each switch experiences less stress. This leads to the possibility of using more efficient switches that have lower voltage and current ratings. Our simulation did kept each H-bridge at a constant duty ratio. The duty cycle for individual H-Bridges would need to be different, or a more complex controller would need to be used in order to evenly apply duty cycles to all the H-bridges. 
    \subsection{NLS Comments}
    \par \hspace{16pt} The benefits that the Nearest Level Switching brings include greater efficiency and the ability to use additional active elements (additional H-bridges) in order to improve output waveform quality \cite{ev-citation-1, Franquelo}. From our analysis it has been shown that by increasing the number of H-bridges and levels, the THD of the output waveform decreases, thus making the output closer to a pure sinusoid. Due to the slow switching nature of the NLS technique, all switches turn on and off only once for the fundamental period of the output waveform, reducing the commutation losses of the switches but increasing the conduction losses. Also, as the number of H-bridges increases, the voltages across each of the switches in the H-bridges decreases (Equation \ref{nls-vthresh}), thus reducing individual switch losses and stresses. At the same time, as the number of H-bridges increases, the number of total switches will increase, thus increasing the total losses as well as the overall complexity of the control for all of the switches \cite{Franquelo}. Thus the optimized number of switches depends on the specifications of the switches used as well as the output voltage and power. As a result, NLS CHB circuits are better suited for some applications more than others \cite{ev-citation-1, Kouro}.
\section{Phase Shifted PWM}
    
    \par \hspace{16pt}
    The Phase Shifted PWM control for a multilevel converter applies a triangular waveform in comparison to a control sinusoidal function in order to obtain the desired PWM for each H-bridge. Each H-bridge's triangle waveform has a phase shift depending on the number of levels:
    \begin{align}
        \theta_{shift} = \frac{360^\circ}{L-1} = \frac{360^\circ}{2N}\label{ps-theta}
    \end{align}
    \par \hspace{16 pt} Where N is the number of H-bridges in the cascade. The DC voltage for each H-bridge level is defined as: 
    \begin{align}
        V_{DC} = \frac{V_{DC,0}}{N} \label{ps-vdc}
    \end{align}
    \par \hspace{16pt} Where $V_{DC,0}$ is the DC voltage required to generate desired AC output voltage in case of a single level inverter.

\section{Phase Shifted PWM Inverter Design}

    \par
\begin{figure}[ht]
        \centering
        \includegraphics[width=0.5\textwidth]{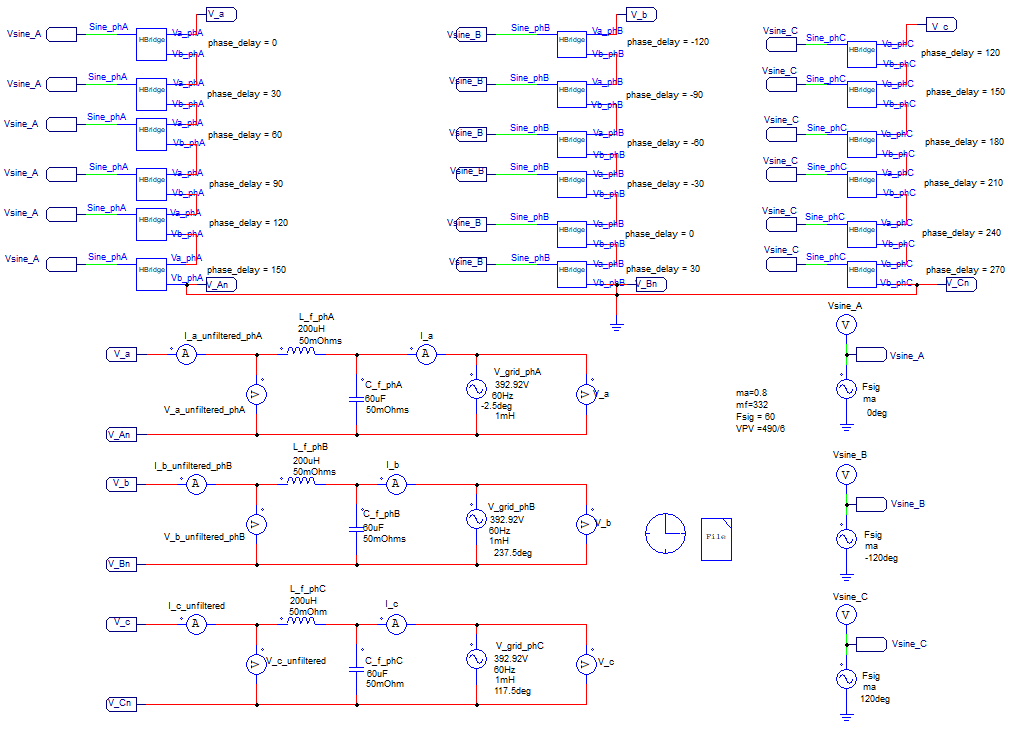}
        \caption{3 phase 6 level cascaded H-bridge inverter with grid as a load}
        \label{ps-3phase}
    \end{figure}

\begin{figure}[ht]
        \centering
        \includegraphics[width=0.5\textwidth]{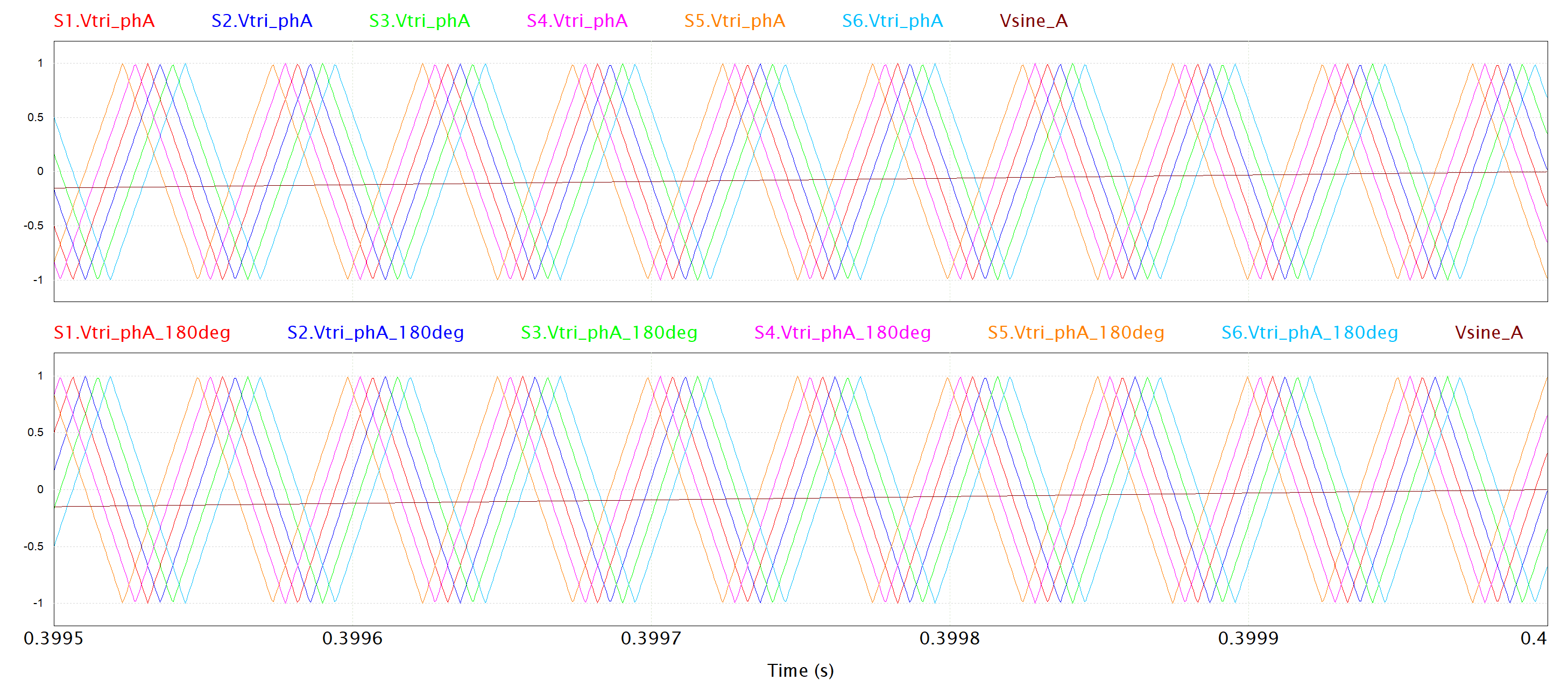}
        \caption{PSPWM carriers (leg A, leg B) for 6 level cascaded H-bridge inverter. Phase shift is $30^\circ$}
        \label{ps-carrier}
    \end{figure}

 \hspace{16pt} For this project, it has been decided to choose cascade consisting of 6 level H-bridge inverters with PSPWM control. The carrier waveforms for all of the 6 levels were triangle waves of 100kHz. The overall 3 phase circuit with load, and level circuitry are shown in Figures \ref{ps-3phase} and \ref{ps-carrier}. Note that the inverter is connected to the grid and the grid has an associated inductance is 1 mH. The grid voltage was given a -2.5° phase shift with respect to inverter output voltage to facilitate current flow from inverter to grid. 
    \\
    \par \hspace{16pt} From our project requirements, the required output voltage was $480 V_{L-L}^{RMS}$ , or $\approx277 V_{L-N}^{RMS}$. From \cite{Mohan} and Equation \ref{ps-vdc}, the total required DC voltage $V_{DC,0}$ and the individual H-bridge DC voltages $V_{DC,level}$ were found as:
    \begin{align*}
        V_{DC,0} = \frac{V_{rms,LL}}{m_a|_{m_a = 0.8}} &* \sqrt{\frac{2}{3}} = \sqrt{\frac{2}{3}} \frac{480}{0.8} \approx 490 V\\
        V_{DC,level} &= \frac{490}{6}\approx81.67V
    \end{align*}
    \par \hspace{16pt} Similarly, from equation \ref{ps-theta}, the individual carrier wave phase shift can be found as:
    \begin{align*}
        \theta_{shift} = \frac{360^\circ}{2*6} = 30^\circ
    \end{align*}
    
    \par \hspace{16pt} Figure \ref{ps-carrier} shows the carrier signals for a 6 level cascaded H-bridge inverter (positive leg control signals are shown in top part, and negative leg control signals are shown in bottom part)
    
    \begin{figure}[ht]
        \centering
        \includegraphics[width=0.5\textwidth]{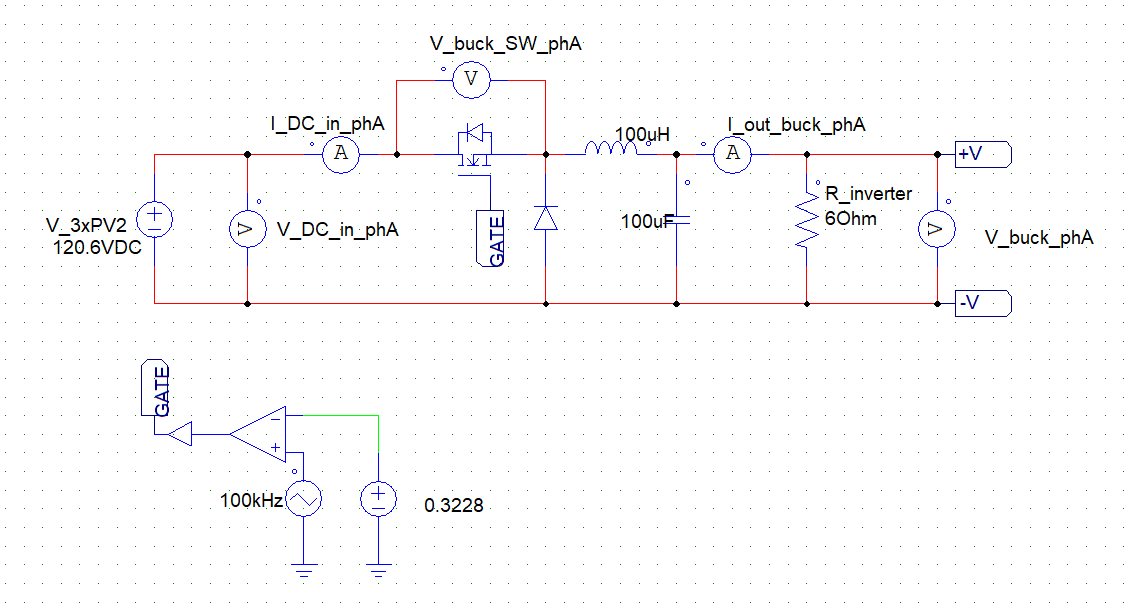}
        \caption{Buck converter required to facilitate 81.67 VDC for inverter level}
        \label{ps-buck}
    \end{figure}
    
    \par \hspace{16pt} Figure \ref{ps-buck} displays the PS-PWM buck converter used. The inductor and capacitor values were calculated using Equations \ref{buck-L} and \ref{buck-C} (Final inductor and capacitor values were chosen to be $100 \mu H$ and $100 \mu F$).This circuit will facilitate power flow from the PV network to the cascaded inverter. Based on the single phase inverter, current draw from the circuit was $\approx13.7 ADC$. That leads to conclude that the inverter's levels will behave as resistive load of ~ $6 \Ohm$ for the buck converter. This figure has been used to design the appropriate buck converter (Equation \ref{f-c-LC}).
    
    \par \hspace{16pt} Based on the data for PV cells in \cite{solar-panels}, it was estimated that two sets of three series PV cells in parallel connected to each buck converter would be needed in order to provide sufficient voltage and current for our inverter. Hence, the input voltage and available current for buck converter are: 120.6 VDC @ 19.42 ADC. This defines duty cycle of buck converter to be ~ 0.677 from Equation \ref{buck-d}.
    
    \begin{figure}[ht]
        \centering
        \includegraphics[width=0.5\textwidth]{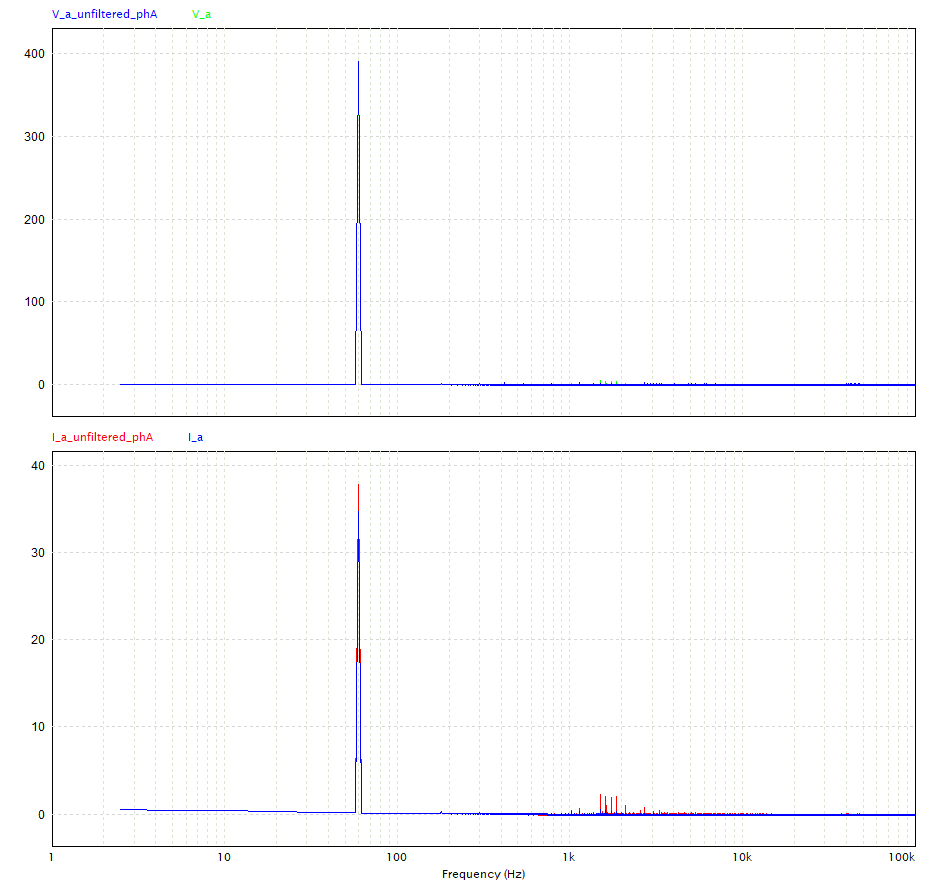}
        \caption{Filtered and unfiltered voltage and current output waveforms}
        \label{ps-frq}
    \end{figure}
    \par \hspace{16pt} Based on the unfiltered voltage and current simulation data, the voltage high frequency harmonics at frequencies above 2 kHz.  Therefore, in order to have a low filtered THD with reasonably high L and C values for filter, it was decided to set cut-off frequency of LC filter at $\approx1453 Hz$. In addition to LC filter, the equivalent grid line inductance of 1 mH also acts as a filter. From Equation \ref{f-c-LC} the calculated filter values were:
    \begin{align*}
        L_f = 200 \mu H\\
        C_f = 60 \mu F 
    \end{align*}
    
\section{PS-PWM Simulation Results}
    \subsection{$3 \phi$ Simulation Data Analysis}
    \par \hspace{16pt} First, the PSIM simulations for a single phase using the default PSIM lossy MOSFET models and lossy reactive elements ($R_{series} = 50m \Omega$) were examined. 
    \begin{figure}[ht]
        \centering
        \includegraphics[width=0.5\textwidth]{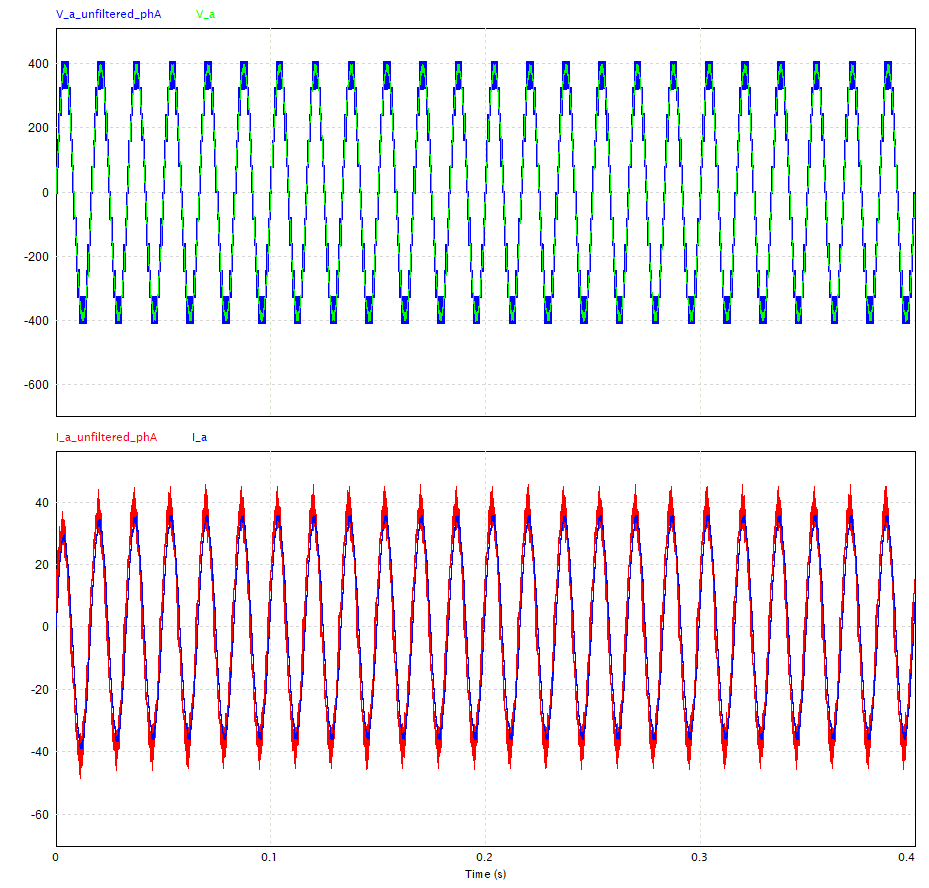}
        \caption{Phase A filtered/unfiltered voltage and current waveforms}
        \label{ps-phasea-wave}
    \end{figure}
    
    \begin{figure}[ht]
        \centering
        \includegraphics[width=0.5\textwidth]{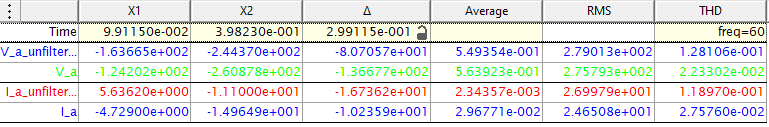}
        \caption{Phase A filtered/unfiltered voltage/current waveform characteristic Figures}
        \label{ps-ph-a-psim}
    \end{figure}
    
    \par \hspace{16pt} Figures \ref{ps-phasea-wave} and \ref{ps-ph-a-psim} display the output voltage and current filtered and unfiltered waveforms as well as the wave characteristics respectively for the single phase lossy simulation. As can be seen, the output voltage met the required line-line voltage of $480 V_{L-L}^{RMS}$. In addition, all of the THD values were below 5\%. The single phase output power was approximately 6.8kW. 
    
    \subsection{$3 \phi$ Simulation Data Analysis}
    \par \hspace{16pt} Next, the PSIM simulations for a three phase circuit using the default PSIM lossy MOSFET models and lossy reactive elements ($R_{series} = 50m \Omega$) were examined.
    \begin{figure}[h]
        \centering
        \includegraphics[width=0.5\textwidth]{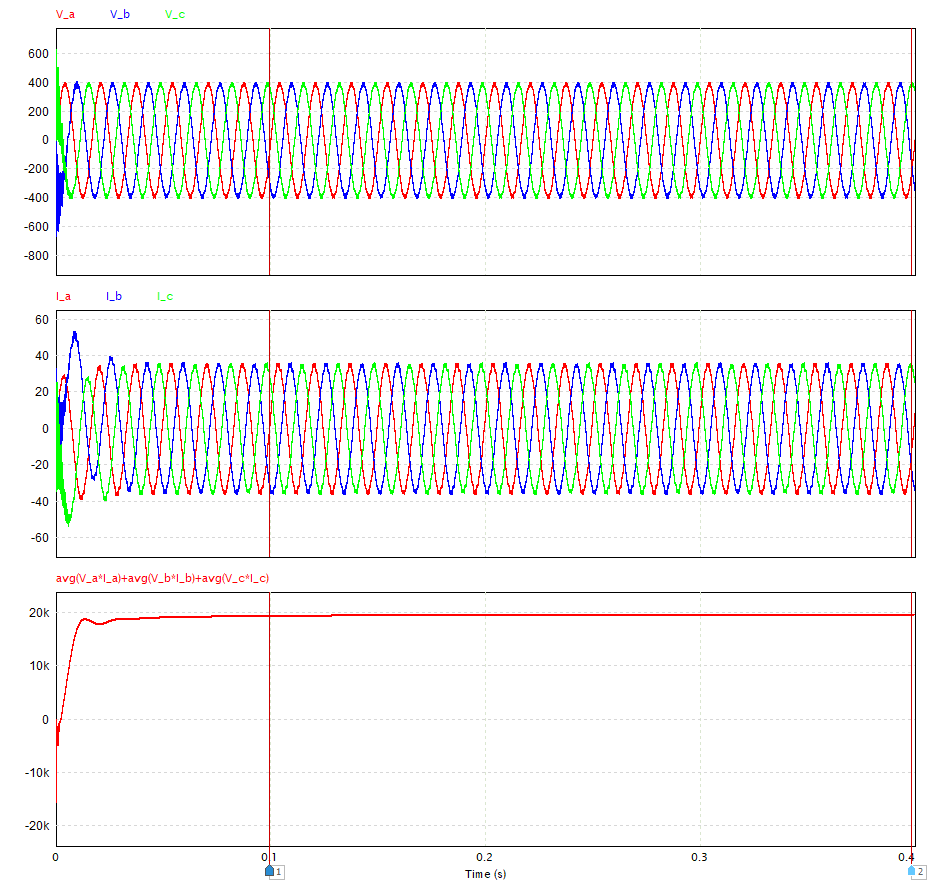}
        \caption{Three phase filtered/unfiltered voltage and current waveforms}
        \label{ps-3phase-wave}
    \end{figure}
    
    \begin{figure}[h]
        \centering
        \includegraphics[width=0.5\textwidth]{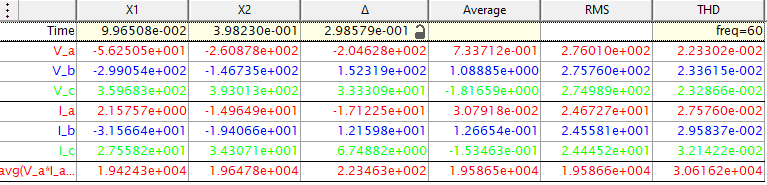}
        \caption{Three phase filtered/unfiltered voltage/current waveform characteristic Figures}
        \label{ps-3phase-psim}
    \end{figure}
    
    \par \hspace{16pt} Figures \ref{ps-3phase-wave} and \ref{ps-3phase-psim} display the output voltage and current filtered and unfiltered waveforms as well as wave their characteristics for the single phase lossy simulation. As can be seen, the output voltage and output current THD values are below the required 5\%. The total output power from the three phase simulation was calculated to be approximately 20.3kW. Thus in order to meet the design requirement of 125kW, at least seven three phase H-bridge inverters of this topology would need to be used. 
    
    \begin{figure}[h]
        \centering
        \includegraphics[width=0.5\textwidth]{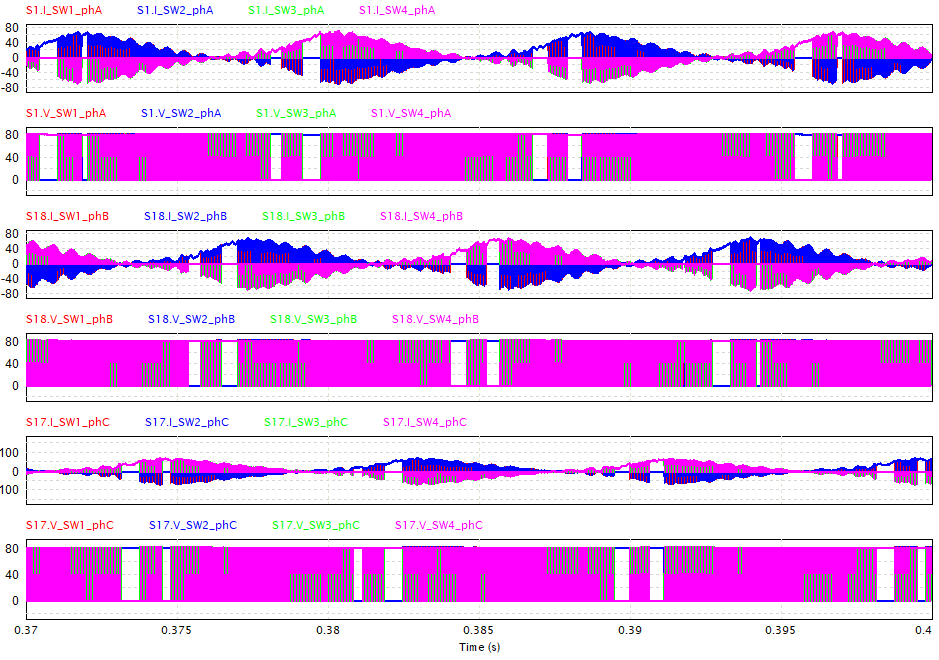}
        \caption{6 level cascaded H-bridge inverter. Switch voltage/current stresses}
        \label{ps-switches}
    \end{figure}
    
    \par \hspace{16pt} Figure \ref{ps-switches} shows the voltages and currents experienced by the H-Bridges of the PS-PWM inverter. Based on the simulation data, the voltage stress on MOSFETs was approximately 85 VDC, and current stress was approximately 40 ADC. Base on design experience, voltage and current ratings are desired to be increased by ~ 150 \% to maintain safe operation at extreme performance cases. That implies that actual switch ratings should be 150 VDC / 60 ADC. 

\section{Conclusions}
        \par \hspace{16pt} Multilevel converters inherently provide desired characteristics for high powered applications; but with them come inherent issues such as more complex structure and operation. The CHB in particular has a structure that allows for very high power applications due to their series connections of isolated power supplies. The drawback of this structure is that the if a single power source is used to supply each of the levels, then the isolation transformer would require currently non-standard transformers with large numbers of secondary windings\cite{Franquelo}. Our solutions, as well as \cite{pv-citation-1, pv-citation-2}, solve this problem by having isolated PV cells. This method assumes that all of the PV's output the same current, but in practice would require more complex control in order to achieve the desired output voltage and power from PV cells that are not providing equal power. In addition, the gate control of each H-bridge would require to be isolated. 
        
        This paper proposes two solutions for creating CHB inverters capable of outputting 125kW at $480 V_{L-L}^{RMS}$. Our simulation results show that such is possible while maintaining a current THD below 5\% as required for IEEE-519 \cite{519}. Multilevel converters such as the CHB have unique features for power quality and modularity. Although they are not commonly used in industry now, they have great potential for the future. 
% \pagebreak

\bibliographystyle{plain}
\bibliography{eece7228.bib}
    \begin{IEEEbiography}
        [{\includegraphics[width=1in,height=1.25in,clip,keepaspectratio]{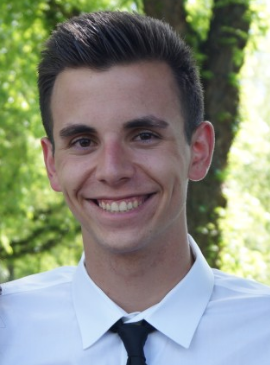}}]{John Buczek}
        currently pursuing a BSMS in Electrical Engineering with concentration in Power Systems at Northeastern University, Boston, MA. His research interests include power electronics, UAVs, and the Wireless Internet of Things.
    \end{IEEEbiography}
    \vfill
    \begin{IEEEbiography}
        [{\includegraphics[width=1in,height=1.25in,clip,keepaspectratio]{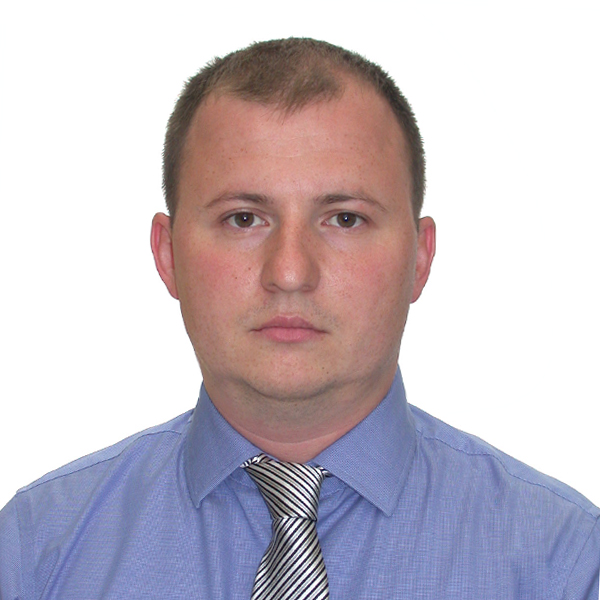}}]{Viktor Ivankevych}
        received B.S (2016) and M.S (2017) degrees in Electrical Engineering from New York University, Brooklyn, NY. He is currently pursuing the Ph.D. degree in Electrical Engineering at Northeastern University, Boston, MA.
His research interests include power electronics and power converters.
    \end{IEEEbiography}
\newpage
\clearpage
    \appendix
        \subsection*{Appendix A: THD Python3 Code}
            \begin{verbatim}
    from math import pi, asin, sqrt

    def rms_value(L):
        '''
        Function to return the rms value for the 'L' level cascade H-bridge inverter
        :param L: int for the number of levels (needs to be odd)
        :retutn: float of the magnitude of rms value
        '''
        N = (L-1)/2                     # Number of H-bridges
        sum_steps = 0                   # summation variable
        for ii in range(0, int(N), 1):  # loop from 0 to N-1
            sum_steps = sum_steps + asin((2*ii+1)/(L-1))*(2*ii + 1)
        return sqrt(1 - (2/(pi*N*N))*sum_steps)
    
    def first_harmonic_rms(L):
        '''
        Funciton to return the rms value of the first harmonic for 
        the 'L' level cascade H-bridge inverter
        :param L: int for the number of levels (needs to be odd)
        :return: float of the magnitude of the rms of the first harmonic
        '''
        N = (L-1)/2                     # Number of H-bridges
        sum_steps = 0                   # summation variable
        for ii in range(0, int(N), 1):  # loop from 0 to N-1
            sum_steps = sum_steps + sqrt(1-((2*ii+1)/(L-1))**2)
        return (8/(pi*1.0*(L-1)))*sum_steps/sqrt(2)
    
    def main():
        '''
        Main Method
        '''
        L_array = range(3, 29, 2)                       # L from 3 to 27 odd integers
        for L in L_array:
            rms = rms_value(L)                          # get rms
            f_harm = first_harmonic_rms(L)              # get first harmonic rms
            thd = sqrt(rms**2 - f_harm**2) / f_harm     # calc THD
            print("Levels: ",L, "Vrms: " ,rms, "V,1,rms: ",f_harm , "THD: ", thd)
    
    if __name__ == "__main__":
        main()
            \end{verbatim}

\end{document}